\documentclass[aps,pre,floatfix]{revtex4}
\usepackage{epsf}
\usepackage{subfigure}
\usepackage{amsmath}
\usepackage{amssymb}
\usepackage{graphicx}
\usepackage{epstopdf}
\begin{document}

\title{\bf Fluctuation theorem for currents and Schnakenberg network theory}

\author{David Andrieux and Pierre Gaspard}
\affiliation{Center for Nonlinear Phenomena and Complex Systems,\\
Universit\'e Libre de Bruxelles, Code Postal 231, Campus Plaine,
B-1050 Brussels, Belgium}

\begin{abstract}
A fluctuation theorem is proved for the macroscopic currents of a system in a 
nonequilibrium steady state,
by using Schnakenberg network theory.
The theorem can be applied, in particular, in reaction systems where
the affinities or thermodynamic forces are defined globally in terms of the cycles
of the graph associated with the stochastic process describing the time evolution.

\vskip 0.1 cm

\end{abstract}

\maketitle

\section{Introduction}
Out-of-equilibrium systems are ubiquitous in Nature and they play an essential role
in many physical, chemical and biological phenomena
\cite{O31,DD36,P67,NP71,N72,MN75,NT77,NP77,JVN84,S76,G90,KP00}.
In an isolated system, a nonequilibrium state will spontaneously relax toward
the state of thermodynamic equilibrium. However, a nonequilibrium steady state
can be maintained in open systems by exchanging energy or matter with
thermostats or chemiostats.
Such reservoirs generate thermodynamic forces -- also called affinities \cite{DD36} --
because they introduce inhomogeneities
in the temperature, the pressure, or the chemical potentials of the
different species of molecules.
In turn, the thermodynamic forces or affinities generate
fluxes of energy or matter across the system,
which contribute to positive entropy production as described
at the macroscopic level by nonequilibrium thermodynamics.

The study of nonequilibrium steady states has a long history.
Already in the XIXth century, nonequilibrium steady states were studied
in linear electric circuits composed of resistors, capacitors, and inductors.
In this context, Kirchhoff developed a network theory to determine
the steady state solution \cite{K1847}.  With the beginning of the XXth century,
the interest shifted to electric circuits with nonlinear elements
such as vacuum tubes and, moreover, fluctuations in electric currents
such as the Nyquist-Johnson and shot noises were discovered, in particular,
under the influence of Einstein's work on Brownian motion and its electric analogue \cite{E05}.
Since 1948, nonlinear elements as transistors and diodes have become dominant
in electric circuits.  These elements have macroscopic 
characteristic curves which are not linear contrary to resistors.
At the mesoscopic level, all the electric circuits present fluctuations in their currents.
In this regard, the currents are random variables and the time evolution of the system
is a stochastic process.  The random variables are described in terms of a
probability distribution.  This latter obeys a master equation ruling the stochastic process.
The random transitions are described in terms of the transition rates 
which may depend on the state variables.

Similar considerations apply to chemical and biochemical reactions.  In the sixties,
Hill and coworkers developed stochastic models 
for the transport of ions across a membrane in biophysics \cite{HK66,H66,H68}.
At the beginning of the seventies, Nicolis and coworkers
developed stochastic models of
autocatalytic chemical reactions to investigate the effects of
fluctuations on the macroscopic nonequilibrium instabilities
\cite{NP71,N72,MN75,NT77,NP77,JVN84}.
Autocatalytic chemical reactions have reaction rates which depend
nonlinearly on the chemical concentrations.
However, we recover a linear description at the mesoscopic level
since the master equation ruling the time evolution 
of the probability distribution is linear.

The analogy between the nonequilibrium steady states of stochastic processes
and linear electric circuits was used since the sixties by Hill \cite{HK66,H66,H68},
Oster and coworkers \cite{OD71,OPK71}, and Schnakenberg \cite{S76}.
Schnakenberg developed a network theory using the methods by Kirchhoff for
linear electric circuits in order to determine the steady state solution \cite{S76}.
In this network theory, a graph is associated with the stochastic process.
The vertices of this graph are the mesoscopic states of the stochastic system and the
edges the possible transitions between the states.
This graph can be decomposed into cycles.
Schnakenberg made the fundamental observation that the ratio
of the products of the transition rates along a cycle and its time reversal
is independent of the mesoscopic states and only depend
on the macroscopic thermodynamic forces or affinities maintaining the
system out of equilibrium \cite{S76}. This observation allowed
Schnakenberg to express several thermodynamic properties
and, in particular, the macroscopic entropy production in terms
of the cycles of network theory \cite{S76}.
Later, the theory of these cycles has been further 
developed by the Qians \cite{QQQ84,JQQ04}.

Recently, remarkable properties of the molecular fluctuations have been discovered
in nonequilibrium systems.  The time evolution of the molecular fluctuations 
can be characterized in terms of their large deviations with respect to their average. 
Several large-deviation dynamical relationships have been obtained such 
as the formulas of the escape-rate and hydrodynamic-mode theories
\cite{GN90,DG95,GD95,TG95,G98,GCGD01}, as well as the
nonequilibrium work theorem \cite{J97,J04} and the fluctuation 
theorems \cite{JQQ04,ECM93,ES94,GC95,G96,R99,K98,C99,LS99,M99,MN03,G04,S05}.
The fluctuation theorems are symmetry relations originating
from microreversibility for the probability distributions of fluctuating quantities,
which are of interest for nonequilibrium thermodynamics and transport properties.
These new large-deviation relationships are under current investigations
to explore the range of their applications.

The purpose of the present paper is to apply the fluctuation theorem
to the nonequilibrium stochastic processes described by Schnakenberg
network theory.  These processes include, in particular, models of
electronic transport in mesoscopic conductors \cite{AWBMJ91}, as well as
chemical and biochemical reactions maintained out of equilibrium by differences
of chemical potentials between the reactants and the products
\cite{NP71,N72,MN75,NT77,NP77,JVN84,HK66,H66,H68}.
Chemical systems are out of equilibrium if the
concentrations of reactants and products are not in their equilibrium ratio
fixed in terms of the standard free enthalpies of formation
of the reactions, as described in chemical thermodynamics \cite{DD36,NP77,S76}.
The differences of free enthalpies between the reactants and products play the
role of thermodynamic forces or affinities.  In this regard,
the thermodynamic forces have energetic and entropic contributions,
and do not reduce to mechanical forces 
as for external electric forces acting on charged particles.
Instead, the macroscopic thermodynamic forces or affinities may differ
from the mesoscopic thermodynamic forces which depend on the mesoscopic states
and are thus time-dependent fluctuating variables as the currents \cite{NP77,JVN84,S76}.
In such systems, the macroscopic thermodynamic forces or affinities should
be defined in terms of cycles, as shown by Schnakenberg \cite{S76}.

Here, we are interested in the fluctuation theorem for the currents which
are fluctuating variables at the mesoscopic level.  Our aim is to give
an algebraic proof of this theorem for the generating function of the currents
associated with Schnakenberg's macroscopic affinities.
For this aim, we have to identify Schnakenberg's cycles.
This is carried out by considering the characteristic determinant of the operator
giving the generating function as its leading eigenvalue.
Our results can be applied to a large variety of processes including nonequilibrium
chemical reactions and electronic transport in mesoscopic conductors \cite{AG04,AG06}.
For non-reactive systems driven in their bulk by an external mechanical force
or for non-reactive systems driven at their boundaries by
time-independent chemical potentials, the fluctuation theorem for the currents
already established in Refs. \cite{G96,LS99,M99} is recovered.

The paper is organized as follows. We introduce the stochastic description in
terms of the master equation in Section \ref{ME-SNT} where
Schnakenberg network theory is summarized. We also illustrate the theory with an example of a chemical reaction network.  In Section \ref{GFT},
the generating function of the currents is introduced by using 
Schnakenberg's graph analysis and the fluctuation theorem 
for the currents is proved in this general framework. In Section \ref{LSFT},
we discuss the connection between 
the fluctuation theorem for the currents and the fluctuation theorem 
for the action functional by Lebowitz and Spohn \cite{LS99}. We also 
show how the fluctuation theorem for the currents can be used
to measure the entropy production in some mesoscopic systems even if 
the microscopic transition rates are not known. 
Conclusions are drawn in Section \ref{Conclusions}.

\section{Master equation and Schnakenberg network theory}
\label{ME-SNT}

\subsection{Master equation and its basic properties}
\label{ME}

Many nonequilibrium phenomena are successfully described at the 
mesoscopic level in terms
of Markovian random processes.  In some simple systems, such 
processes can even be exactly derived from the underlying 
deterministic dynamics \cite{TG95,G98} by introducing an appropriate 
partition of the phase space.  In other systems, such processes can 
be rigorously derived from the underlying Hamiltonian classical or 
quantum dynamics in some scaling limit \cite{S80}.
Continuous-time random processes are ruled by an evolution equation, 
called the master equation,
for the probability to find the system in a coarse-grained state 
$\omega$ at time $t$:
\begin{equation}
\frac{d P(\omega,t)}{dt} = \sum_{\rho ,\omega'} \big[ 
W_{\rho}(\omega'\vert \omega)P(\omega',t)
- W_{\rho}(\omega \vert \omega')P(\omega,t) \big]
\label{master.equ.}
\end{equation}
The quantities $W_{\rho}(\omega'\vert \omega)$ denote the rates 
of the transitions
$\omega'{\overset{\rho}{\to}}\omega$
allowed by the elementary processes $\rho=1, 2,..., r$.
Each one of these elementary processes may independently contribute 
to the entropy production so that it is important to separate them in
the master equation.  This is the case in nonequilibrium reaction systems
where each reaction $\rho$ contributes to the entropy production 
\cite{JVN84,S76,G04,JQQ04,AG04}.

The master equation is known to obey a $H$-theorem for the entropy
\begin{equation}
S(t)= \sum_{\omega} S^0(\omega) P(\omega,t) - \sum_{\omega} 
P(\omega,t) \ln P(\omega,t)
\label{entropy}
\end{equation}
associated with the probability distribution $P(\omega,t)$ 
describing the state of the system at the time $t$ \cite{JVN84,S76}. 
$S^0(\omega)$ denotes the entropy due to the statistical distribution of all the degrees of freedom which are not specified by the coarse-grained state $\omega$ \cite{G04}. 
For instance, if the coarse-grained state $\omega$ only specifies the numbers of the particles of the different species, $S^0(\omega)$ is the entropy of the statistical distribution of the
positions and momenta of the particles.  
The second term is the contribution to entropy due to the statistical distribution $P(\omega,t)$ 
of the numbers of particles \cite{G04}.  
The entropy is here calculated
in the units of Boltzmann's constant $k_{\rm B}\simeq 1.38 \; 10^{-23}$ J/K.
The time derivative of the entropy is given by
\begin{equation}
\frac{dS}{dt} = \frac{d_{\rm e}S}{dt}+\frac{d_{\rm i}S}{dt}
\end{equation}
in terms of the entropy flux $d_{\rm e}S/dt$ and the non-negative 
entropy production
\begin{equation}
\frac{d_{\rm i}S}{dt} =
\frac{1}{2} \sum_{\rho ,\omega,\omega'} J_{\rho}(\omega,\omega') 
A_{\rho}(\omega,\omega') \geq 0
\label{prod.entropie}
\end{equation}
where
\begin{equation}
J_{\rho}(\omega, \omega ') \equiv P(\omega,t) W_{\rho}(\omega \vert \omega')
- P(\omega',t) W_{\rho}(\omega' \vert \omega)
\label{meso.fluxes}
\end{equation}
is the current of the transition $\omega{\overset{\rho}{\to}}\omega'$ and
\begin{equation}
A_{\rho}(\omega,\omega') \equiv \ln 
\frac{P(\omega,t)W_{\rho}(\omega\vert \omega')}
{P(\omega',t)W_{\rho}(\omega'\vert  \omega)}
\label{meso.affinities}
\end{equation}
the corresponding affinity \cite{JVN84,S76,G04}.  The $H$-theorem asserts that the
entropy production (\ref{prod.entropie}) is always non-negative 
in agreement with the second law of thermodynamics 
and thus characterizes the irreversibility of the process.

In a stationary state where $d P/dt=0$, the master equation (\ref{master.equ.})
can also be written as
\begin{equation}
\sum_{\rho ,\omega '} J_{\rho}(\omega' ,\omega) = 0
\label{Kirch}
\end{equation}
in terms of the currents (\ref{meso.fluxes}),
which is nothing else than the Kirchhoff current law \cite{S76}.

In the equilibrium stationary state, the conditions of detailed balance
\begin{equation}
P_{\rm eq}(\omega)W_{\rho}(\omega\vert \omega')
= P_{\rm eq}(\omega')W_{\rho}(\omega'\vert  \omega)
\label{DB}
\end{equation}
are satisfied for all the possible forward and backward transitions
$\omega{\overset{\rho}{\rightleftharpoons}}\omega'$.
Consequently, the currents (\ref{meso.fluxes})
and the affinities (\ref{meso.affinities})
as well as the entropy production (\ref{prod.entropie}) all vanish at equilibrium.


\subsection{Schnakenberg network theory}
\label{SNT}

Schnakenberg has shown that many of the fundamental properties of a
nonequilibrium random process can be investigated and understood
by carrying out the analysis of the graph associated with the master 
equation \cite{S76}.
Moreover, Schnakenberg found that the nonequilibrium constraints of a system
are related to the affinities of the cycles of the graph, obtaining
in this way an early form of the fluctuation theorem \cite{S76}.

Let us first introduce the tools needed for the graph analysis of the 
random process.
For a system ruled by the master equation (\ref{master.equ.}), 
a graph $G$ is associated as follows \cite{S76}:
each state $\omega$ of the system corresponds to a vertex
while the edges represent the different transitions
$\omega{\overset{\rho}{\rightleftharpoons}}\omega'$
allowed between the states.
Accordingly, two states can be connected
by several edges if several elementary processes $\rho$ allow 
transitions between them.

A method has been provided by Schnakenberg to identify all the cycles of a graph \cite{S76}.
This method is based on the definition of a maximal tree $T(G)$ of 
the graph $G$, which should satisfy the following properties:

(1) $T(G)$ is a covering subgraph of $G$, i.e., $T(G)$ contains all 
the vertices of $G$ and all the edges of $T(G)$ are edges of $G$;

(2) $T(G)$ is connected;

(3) $T(G)$ contains no circuit, i.e., no cyclic sequence of edges.

In general a given graph $G$ has several maximal trees $T(G)$.

The edges $l$ of $G$ which do not belong to $T(G)$ are called the chords of $T(G)$.
If we add to $T(G)$ one of its chords $l$, the resulting subgraph $T(G)+l$
contains exactly one circuit, ${\cal C}_l$, which is obtained from $T(G)+l$ by removing
all the edges which are not part of the circuit.  
The set of circuits $\{{\cal C}_1,{\cal C}_2,...,{\cal C}_l,...\}$ is called a fundamental set. 
An arbitrary orientation can be assigned to each circuit ${\cal C}_l$
to define a cycle ${C}_l$. A maximal tree $T(G)$ together with 
its associated fundamental set of cycles $\{{C}_1,{C}_2,...,{C}_l,...\}$ provides
a decomposition of the graph $G$.  We notice that the maximal tree $T(G)$
can be chosen arbitrarily because each cycle ${C}_l$ can be redefined by
linear combinations of the cycles of the fundamental set.

An orientation is given to each edge of the graph $G$. The directed edges are thus defined by
\begin{equation}
e \equiv \omega \overset{\rho}{\rightarrow} \omega'
\label{directed.edge}
\end{equation}
Let $f$ be a directed subgraph of $G$.
The orientation of the subgraph $f$ 
with respect to its edges $\{ e\}$ is described by introducing the 
quantity
\begin{equation}
S_e(f) \equiv \begin{cases} +1 & \text{if $e$ and $f$ are parallel,} \\
                -1 & \text{if $e$ and $f$ are antiparallel,} \\
                0 & \text{if $e$ is not in $f$}.
                \end{cases}
\label{orientation}
\end{equation}
where $e$ and $f$ are said to be parallel (resp. antiparallel)
if $f$ contains the edge $e$ in its reference (resp. opposite) 
orientation.
It will be useful for the following to write the affinities and the 
fluxes in terms of these numbers.
In the nonequilibrium steady state, the affinity of the edge (\ref{directed.edge}) is defined by
\begin{equation}
A_e \equiv \ln  \frac{P_{\rm st}(\omega)W_{\rho}(\omega\vert 
\omega')}{P_{\rm st}(\omega')W_{\rho}(\omega'\vert  \omega)}
\end{equation}
The affinity of an arbitrary cycle ${C}$ is then written as
\begin{equation}
{\cal A}({C}) = \sum_e S_e({C})A_e
\end{equation}
where the sum is taken over all the edges $e$ of the graph $G$. It 
can be equally written as
\begin{equation}
{\cal A}({C}) = \sum_e S_e({C})B_e
\label{A.decomp}
\end{equation}
with
\begin{equation}
B_e \equiv \ln \frac{W_{\rho}(\omega\vert 
\omega')}{W_{\rho}(\omega'\vert \omega)}
\label{Be}
\end{equation}
Indeed, the terms $\ln \left[ P_{\rm st}(\omega )/P_{\rm st}(\omega ')\right]$
exactly cancel each other when calculated on a closed path.
Moreover, we can introduce a scalar product
\begin{equation}
({C},{C}_l) = S_l({C}) S_l({C}_l)
\label{PS}
\end{equation}
where ${C}_l$ is the graph of the fundamental set
associated with the chord $l$ as introduced above.
It takes the value $+1$ ($-1$) if ${C}$ contains the chord $l$
with the same (opposite) orientation, and $0$ otherwise.
The affinity of any cycle ${C}$ can then be decomposed as
\begin{equation}
{\cal A}({C}) = \sum_l \ ({C},{C}_l) {\cal A}({C}_l)
\label{Adecom}
\end{equation}
where the sum extends over all the chords.
This shows in particular that the affinity of an arbitrary cycle
is a linear combination of the affinities of a fundamental set \cite{S76}.
Finally, applying successively Kirchhoff current law (\ref{Kirch}),
the mean current traversing the edge $e$ is written as:
\begin{equation}
J_e = \sum_l S_e({C}_l) J({C}_l)
\label{KCL.S}
\end{equation}

Nonequilibrium constraints are imposed to a system if the temperature, the pressure, or the chemical potentials differ between the reservoirs surrounding the system.  These constraints are characterized by
the global thermodynamic forces or affinities defined by the differences of the temperatures, pressures, or chemical potentials of the reservoirs.  These global affinities do not directly appear in the transition rates
$W_{\rho}(\omega\vert\omega')$ of the master equation (\ref{master.equ.}).  These rates depend
on the temperature, the pressure, or the chemical potentials of the reservoir responsible for the transition $\omega \overset{\rho}{\rightarrow} \omega'$ and also on the states $\omega$ and $\omega'$, although the global affinities $\{ {\cal A}_{\gamma}\}$ are defined by the {\it differences} of the temperatures, pressures, or chemical potentials of the reservoirs and are independent of the particular states $\omega$ or $\omega'$.
In this regard, the global affinities are macroscopic.

As observed by Schnakenberg \cite{S76}, the global macroscopic affinities $\{ {\cal A}_{\gamma}\}$ 
can be identified by considering the cycles $\{{C}_l\}$ of the graph $G$ associated with the stochastic process. The fact is that the ratio of the products
of the transition rates along the two possible time directions of any cycle ${C}_l$
of the graph is independent of the states composing the cycle and only depends
on the global macroscopic affinities $\{ {\cal A}_{\gamma}\}$:
\begin{equation}
\prod_{e \in {C}_l}
\frac{W_{\rho}(\omega\vert\omega')}{W_{\rho}(\omega'\vert\omega)}
= {\rm e}^{{\cal A}({C}_l)} = {\rm e}^{{\cal A}_{\gamma}}
\label{ratio}
\end{equation}
where $e \in {C}_l$ denotes the edges (\ref{directed.edge}) in the cycle ${C}_l$.
The conditions (\ref{ratio}) are satisfied in a large class of processes including diffusion processes in lattice gases, nonequilibrium chemical reactions, and electronic transport in mesoscopic conductors \cite{S76,LS99,AG04,AG06}.
We notice that there can exist more cycles ${C}_l$
than currents $\gamma$ between the reservoirs.  The reason is that 
the graph describes
all the possible states and transitions at the mesoscopic level while 
the currents $\gamma$
between the reservoirs are typically macroscopic and fewer than the 
mesoscopic states.
The affinities or thermodynamic forces ${\cal A}({C}_l)$
associated with the various cycles ${C}_l$ of a graph $G$ may 
thus take the same value for all cycles corresponding to the same 
current $\gamma$
between two given reservoirs: ${\cal A}({C}_l) = {\cal A}_\gamma$ 
for all ${C}_l \in \gamma$.
The present paper is devoted to the class of systems where the thermodynamic forces or affinities
are defined by the Schnakenberg conditions (\ref{ratio}).  These conditions are weaker than in systems
with external mechanical forces where the affinities can be directly identified at the level
of the transition rates $W_{\rho}(\omega\vert\omega')$ themselves.
Here, the affinities only appear after closing the cycles ${C}_l$, which significantly complicates the reasoning.

\subsection{Example: Schl\"ogl's reaction scheme}
\label{Example}

\begin{figure}[h]
\centerline{\scalebox{1.1}{\includegraphics{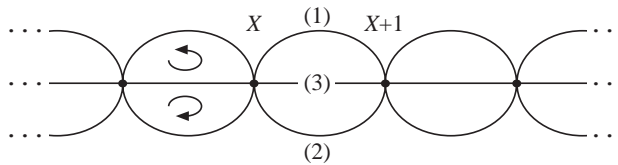}}}
\caption{Graph $G$ associated with the reaction network (\ref{react.1})-(\ref{react.3}).}
\label{fig1} 
\end{figure}

Our purpose in this section is to illustrate Schnakenberg's network theory with the example of the nonlinear chemical network
\begin{eqnarray}
\mathrm{A} &\overset{k_{1}}{\underset{k_{-1}}{\rightleftharpoons}}& \mathrm{X} \label{react.1} \\
\mathrm{B} + \mathrm{X} &\overset{k_{2}}{\underset{k_{-2}}{\rightleftharpoons}}& \mathrm{2X} \label{react.2} \\
\mathrm{C} +\mathrm{2X} &\overset{k_{3}}{\underset{k_{-3}}{\rightleftharpoons}}& \mathrm{3X}
\label{react.3}
\end{eqnarray}
Particles enter into the system from three different reservoirs, 
$\mathrm{A}$, $\mathrm{B}$, and $\mathrm{C}$.  The particles of the species X 
are produced by three different reactions from so many reactants. 
The reaction constants are denoted by $k_{\rho}$ with $\rho=1,2,3$. Out of equilibrium, fluxes of matter will cross the system between the three reservoirs. 
In a stationary state only two of them will be independent and it is necessary to distinguish them as they are independent macroscopic observables. 
According to the mass-action law \cite{NP71,N72,MN75,NT77,NP77,JVN84,S76,G90,G04,AG04}, 
the transition rates of these reactions are proportional to the concentrations and given by
\begin{eqnarray} 
W_{1}(X\vert X+1) &=& k_{1} [\mathrm{A}] \Omega  \\ 
W_{1}(X\vert X-1) &=& k_{-1} X \\
W_{2}(X\vert X+1) &=& k_{2} [\mathrm{B}] X \\ 
W_{2}(X\vert X-1) &=& k_{-2} X \frac{X-1}{\Omega} \\
W_{3}(X\vert X+1) &=& k_{3} [\mathrm{C}] X \frac{X-1}{\Omega} \\ 
W_{3}(X\vert X-1) &=& k_{-3} X \frac{X-1}{\Omega}\frac{X-2}{\Omega}
\end{eqnarray} 
where $X$ is the number of particles of the species X, $[\cdot]$ denotes the mean concentration in the reservoirs, and $\Omega$ is the volume of the system. 
The transition rates are nonlinear functions of the internal state $X$. 
The state of the system is described in terms of the probability $P(X,t)$ that the system contains
the number $X$ of particles of the species X at the time $t$.
The master equation ruling the time evolution takes the form:
\begin{eqnarray}
\frac{d}{dt}P(X,t) = \sum_{\rho=1}^3 \sum_{\nu=\pm 1} \Big[ W_{\rho}(X-\nu \vert X) P(X-\nu,t)
- W_{\rho}(X \vert X-\nu) P(X,t) \Big]
\label{master.eq}
\end{eqnarray}
We notice that neither the macroscopic currents nor the macroscopic
affinities are apparent in the master equation
contrary to the systems where the affinities are given in terms of mechanical forces.
In order to identify them, we use the graph analysis of Schnakenberg \cite{S76}.
The graph of this stochastic process is depicted in Fig. \ref{fig1}.

At equilibrium, the detailed balance conditions (\ref{DB}) hold which imply that
\begin{equation}
\frac{k_1[\mathrm{A}]}{k_{-1}} = \frac{k_2[\mathrm{B}]}{k_{-2}} = \frac{k_3[\mathrm{C}]}{k_{-3}} = \frac{\langle X\rangle_{\rm eq}}{\Omega}
\label{DB.chem}
\end{equation}
and the equilibrium probability distribution $P_{\rm eq}(X)$ is Poissonian of average $\langle X\rangle_{\rm eq}$.  Out of equilibrium, the detailed balance conditions are no longer satisfied and the probability distribution $P_{\rm st}(X)$ of the nonequilibrium steady state is known to be nonPoissonian for such nonlinear chemical networks \cite{NP71,N72,MN75,NT77,NP77}.
For the present model, the steady state is given by: 
\begin{eqnarray}
P_{\rm st} (X) &=& P_{\rm st} (0) \prod_{x=1}^{X} \frac{\sum_{\rho} W_{\rho}(x -1 \vert x)}{\sum_{\rho}W_{\rho}(x \vert x-1)}
 \nonumber \\
&=& P_{\rm st} (0) \prod_{x=1}^{X} \frac{k_{1} [\mathrm{A}] \Omega+k_{2} [\mathrm{B}] (x-1)+k_{3} [\mathrm{C}](x-1)(x-2)\Omega^{-1}}{k_{-1} x+k_{-2}x (x-1)\Omega^{-1} +k_{-3} x (x-1) (x-2)\Omega^{-2}}
\label{st.prob}
\end{eqnarray}

An affinity can be associated with each reaction ($\rho=1,2,3$) according to Eq. (\ref{meso.affinities}):
\begin{equation}
A_{\rho} (X) \equiv  \ln \frac{P_{\rm st} (X) W_{\rho}(X\vert X+1)}{P_{\rm st} (X+1) W_{\rho}(X+1\vert X)} 
\end{equation}
These mesoscopic affinities depend on the internal state $X$, and thus fluctuates in time.
Therefore, they do not correspond to the time-independent macroscopic affinities.
These latter are instead given by the Schnakenberg conditions (\ref{ratio}). A possible maximal tree is given by choosing all edges corresponding to reaction $\rho=3$. The chords correspond in this case to all edges associated with the reactions $\rho=1,2$. The corresponding cycles are depicted in Fig. \ref{fig1}. They start from the state $X$
and go to the state $X+1$ by the edge $\rho=3$ and return to the state $X$
by the edges $\rho=1$ or $\rho=2$. The corresponding macroscopic affinities are given by
\begin{equation}
{\cal A}_{\rho} \equiv \ln \frac{W_{3}(X\vert X+1)W_{\rho}(X+1\vert X)}{W_{3}(X+1\vert X)W_{\rho}(X\vert X+1)}
\label{affinities.react}
\end{equation}
or
\begin{eqnarray}
{\cal A}_{1} &=& \ln \frac{k_{-1}k_{3} [\mathrm{C}]}{k_{-3} k_{1} [\mathrm{A}]} , \quad {\cal A}_{2} = \ln \frac{k_{-2}k_{3} [\mathrm{C}]}{k_{-3} k_{2} [\mathrm{B}]}
\label{af.schlogl}
\end{eqnarray}
which are independent of the internal state $X$ and thus constant in time as it should. 

This example shows that the mesoscopic affinities (\ref{meso.affinities}) depend in general on the state
of the system and will thus fluctuate in time together with the corresponding currents (\ref{meso.fluxes})
along a stochastic trajectory of the process.  This problem is overcome by introducing
the macroscopic affinities defined by the Schnakenberg conditions (\ref{ratio}).
These latter no longer depend on the mesoscopic state and are thus independent
of time as expected for nonequilibrium constraints from macroscopic reservoirs.

\section{The fluctuation theorem for the currents}
\label{GFT}

We derive in this section a fluctuation theorem for the currents for the class of systems defined 
by the Schnakenberg conditions (\ref{ratio}).
We give a different demonstration than in reference \cite{AG04}.
Here, we adopt a definition of the currents based on a given maximal tree.
This definition allows us more freedom than in reference \cite{AG04}
because the choice of the maximal tree
can be used to derive several large-deviation relations in a given system.
Moreover, we here consider general processes, not only chemical reactions.
Finally, the present proof provides an efficient numerical procedure
to compute the quantities of interest.
The connection with the fluctuation theorem for the Lebowitz-Spohn action functional \cite{LS99}
is discussed at the end of this section.

We now begin the derivation of the fluctuation theorem for the currents.
The instantaneous current on the chord 
$l$ is defined by
\begin{equation}
j_l (t) \equiv \sum_{n=-\infty}^{+\infty} S_{l}(e_n)\, \delta (t-t_n)
\label{j_l}
\end{equation}
where $t_n$ is the time of the random transition $e_n$ during a path of the stochastic process.
We use the convention that $j_l$ is oriented as the graph $G$
since $S_{l}(e_n)$ is equal to $(-)1$ if the transition $e_n$ is 
(anti)parallel to the chord $l$. In particular, the fundamental cycles are such that $S_l({C}_l)=1$.
This means that the cycles are oriented in the same way as the chords $l$ but this is not restrictive.
The current (\ref{j_l}) is a fluctuating random variable.
The so-called Helfand moment \cite{H60}
associated with the current is defined as the integrated current by
\begin{equation}
G_l(t) \equiv \int_0^{t} dt' j_l (t')
\label{Helfand}
\end{equation}

The quantity of interest is the generating function
of the \textit{independent} currents crossing the aforementioned 
chords of the system
\begin{equation}
Q(\{\lambda_l \}; \{ {\cal A}_l \}) \equiv
\lim_{t\to\infty} -\frac{1}{t} \ln \langle {\rm e}^{- \sum_l 
\lambda_l \int_0^{t} dt' j_l (t')}\rangle
\label{Q.flux}
\end{equation}
where the sum is taken over all the chords and
\begin{equation}
{\cal A}_l \equiv {\cal A}({C}_l)
\end{equation}
We notice that the generating function (\ref{Q.flux}) can be written 
as \cite{GD95}
\begin{equation}
Q(\{\lambda_l \}; \{ {\cal A}_l \}) \equiv
\lim_{t\to\infty} -\frac{1}{t} \ln \langle {\rm e}^{- \sum_l 
\lambda_l G_l(t)}\rangle
\label{Q.flux.G}
\end{equation}
in terms of the Helfand moments (\ref{Helfand}).

We shall first prove the fluctuation theorem for the currents
(\ref{j_l}), according to which
the generating function of the fluctuating quantities (\ref{Helfand})
obeys the symmetry
\begin{equation}
Q(\{\lambda_l \}; \{ {\cal A}_l \}) = Q(\{ {\cal A}_l - \lambda_l \}; 
\{ {\cal A}_l \})
\label{CFT.Q}
\end{equation}
under the Schnakenberg conditions (\ref{ratio}).

In order to prove Eq. (\ref{CFT.Q}), we define
\begin{equation}
F_t(\omega,\{\lambda_l\}) \equiv \langle {\rm e}^{- \sum_l \lambda_l G_l 
(t)}\rangle_{\omega}
\end{equation}
as the mean value of $\ \exp \left[- \sum_l \lambda_l G_l (t)\right]$
with the initial condition that the system is in state $\omega$ at time $t=0$.
We want to derive an evolution equation for this quantity. The 
quantity $F_{t+dt}(\omega,\{\lambda_l\})$ can be calculated by 
considering all the possible
transitions occurring in an infinitesimal time $dt$ starting from state $\omega$ at time $t=0$, 
before evolving the ensemble average for a time $t$ starting from the state $\omega'$ reached at time $dt$.
All these possible transitions must be weighted with their respective 
occurrence probabilities
multiplied by the contributions of the exponential factors
$\exp \left[- \sum_l \lambda_l G_l (t)\right]$.
The quantity $F_{t+dt}(\omega,\{\lambda_l\})$ can then be written as
\begin{eqnarray}
F_{t+dt}(\omega,\{\lambda_l\}) = \sum_{\rho,\omega '}P_{dt}(\omega 
\overset{\rho}{\rightarrow} \omega') \exp \left[- \sum_l S_l (\omega 
\overset{\rho}{\rightarrow} \omega') \lambda_l \right] \ 
F_{t}(\omega',\{\lambda_l\})
\label{Gtdt}
\end{eqnarray}
in terms of the orientation numbers (\ref{orientation}) with a sum taken over all the chords.
In Eq. (\ref{Gtdt}), the exponential factor thus comes from the contributions of the quantity
$\exp \left[- \sum_l \lambda_l G_l (t)\right]$
during the transition $\omega \overset{\rho}{\rightarrow} \omega'$.
To simplify the notations, we introduce the quantities:
\begin{equation}
{\rm e}^{- z_{\rho}(\omega \vert \omega')} \equiv {\rm e}^{- \sum_l S_l 
(\omega \overset{\rho}{\rightarrow} \omega') \lambda_l } = \begin{cases} \exp 
(-\lambda_l) & \text{if $\omega \overset{\rho}{\rightarrow} \omega'$ 
corresponds to the chord $l$ in its reference orientation} \\
\exp (+\lambda_l) & \text{if $\omega \overset{\rho}{\rightarrow} 
\omega'$ corresponds to the chord $l$ in the opposite orientation} 
\\
1 & \text{otherwise}
\end{cases}
\end{equation}
which depend on whether a transition is taken along one of the chords 
$\{ l \}$ or not.
$P_{dt}(\omega \overset{\rho}{\rightarrow} \omega')$
corresponds to the probability to jump to the state $\omega'$ in a 
time $dt$ along the transition $\rho$ and starting from $\omega$. 
This conditional probability is given by
\begin{equation}
P_{dt}(\omega \overset{\rho}{\rightarrow} \omega') = \begin{cases} 
W_{\rho}(\omega\vert\omega') dt + O(dt^2) & \text{if $\omega \not= 
\omega'$} \\
1- \sum_{\rho,\omega'} \ W_{\rho}(\omega\vert\omega') dt + O(dt^2) & 
\text{if $\omega = \omega'$}
\end{cases}
\end{equation}
Inserting this formula into Eq. (\ref{Gtdt}), we get
\begin{eqnarray}
F_{t+dt}(\omega,\{\lambda_l\}) &=& \sum_{\rho,\omega ' \not= 
\omega}dt \ W_{\rho}(\omega \vert \omega') \ {\rm e}^{- 
z_{\rho}(\omega\vert \omega')} F_{t}(\omega',\{\lambda_l\}) \nonumber 
\\
&+& \left[1 - dt \sum_{\rho,\omega'} \ 
W_{\rho}(\omega\vert\omega')\right] F_{t}(\omega,\{\lambda_l\}) + 
O(dt^{2})
\end{eqnarray}
so that the time derivative of $F_{t}(\omega,\{\lambda_l\})$ is given by
\begin{eqnarray}
\frac{d}{dt} F_{t}(\omega,\{\lambda_l\}) &=& \sum_{\rho,\omega'} 
\left\{ W_{\rho}(\omega\vert \omega') \exp \left[- \sum_l S_l (\omega 
\overset{\rho}{\rightarrow} \omega') \lambda_l \right] 
F_{t}(\omega',\{\lambda_l\}) - W_{\rho}(\omega\vert\omega')
F_{t}(\omega,\{\lambda_l\}) \right\} \nonumber \\
&\equiv& \hat L_{\{\lambda_l\}} \ F_{t}(\omega,\{\lambda_l\})
\label{G.eq}
\end{eqnarray}
with the initial condition $F_{t=0}(\omega,\{\lambda_l\}) = 1$.
Whereupon we find that
\begin{equation}
\langle {\rm e}^{- \sum_l \lambda_l \int_0^{t} dt' j_l (t')}\rangle
= \sum_{\omega } P_{\rm st} (\omega ) F_{t}(\omega,\{\lambda_l\})
= \sum_{\omega ,\omega'} P_{\rm st} (\omega )\left[{\rm e}^{\hat 
L_{\{\lambda_l\}}t}\right]_{\omega ,\omega'}
\end{equation}
With our conditions on the transition rates, the Perron-Frobenius 
theorem is of application and there is a unique maximal eigenvector 
$V_{\{\lambda_l\}}$ (which is positive)
\begin{equation}
\hat L_{\{\lambda_l\}}V_{\{\lambda_l\}} = -Q(\{\lambda_l\})\, V_{\{\lambda_l\}}
\end{equation}
so that the limit in Eq. (\ref{Q.flux}) exists and the leading 
eigenvalue of the operator (\ref{G.eq})
gives the generating function of the currents.

We continue our analysis by looking directly at the eigenvalues given 
by the roots of the characteristic equation of the operator (\ref{G.eq}). We will make the 
connection between the Schnakenberg graphical analysis and this 
equation, showing that the characteristic equation presents the 
symmetry (\ref{CFT.Q}) of the fluctuation theorem
for the currents. Consequently, the eigenvalues of the operator 
(\ref{G.eq}) should also have the
symmetry. However, the eigenvectors do not necessarily present the 
symmetry so that
the long-time limit must be performed in order to obtain the leading 
eigenvalue and that the symmetry arises in the generating function 
according to the Perron-Frobenius theorem.

The eigenvalues of an operator $\hat L$ acting on a finite state 
space are given by the roots
of its characteristic determinant:
\begin{equation}
\det\left(\hat L-s \, \hat I\right) = 0
\label{ch.det}
\end{equation}
Here, we introduce the matrix $\hat M$ composed of the elements
\begin{equation}
M_{ij} \equiv \begin{cases} L_{ii}-s & \text{if $i=j$}\\
L_{ij} & \text{if $i\neq j$}
\end{cases}
\end{equation}
The characteristic determinant (\ref{ch.det}) is thus given by the 
following expression
\begin{equation}
\det\hat M = \sum_{p \in {\rm Sym}(n)} \ (-1)^{\pi(p)} 
\ M_{1p(1)} M_{2p(2)} \cdots M_{np(n)}
\label{det}
\end{equation}
where the sum is taken over all the permutations of the symmetric 
group of $n$ objects, ${\rm Sym}(n)$. This group contains $n!$ 
elements. $\pi(p)$ is the parity of the permutation $p$ and takes the 
value $\pm 1$ depending on whether the permutation can be decomposed 
into an even or odd number of transpositions.

Each term in the determinant (\ref{det}) is thus a product of the 
coefficients of the linear equation (\ref{G.eq}). A graphical 
representation of each term can be obtained by selecting for each 
vertex an edge (with the additional possibility to stay at this 
vertex). If an edge connecting
two states is not present, the corresponding terms in the determinant 
will of course be absent.

There is the additional constraint that there cannot be two edges 
pointing towards the same vertex.
This results from the fact that the permutations form a group and are 
therefore inversible.
This constraint implies that the determinant only contains terms 
corresponding to
some cycles of the graph plus some fixed vertex (called full circuit 
in graph theory).
A simple example is the case of a $3 \times 3$ matrix $\hat A$ where 
the determinant contains the term $a_{11}a_{22}a_{33}$ corresponding 
to three fixed vertices, the terms $a_{11}a_{23}a_{32}$, 
$a_{13}a_{22}a_{31}$, $a_{12}a_{21}a_{33}$ where there is one fixed 
vertex and a cycle of length $2$, and finally the terms 
$a_{12}a_{23}a_{31}$, $a_{13}a_{21}a_{32} $ with no fixed vertex and 
a cycle of length $3$. For general matrices of $n\times n$, this 
property results from the representation of the permutations $p \in 
{\rm Sym}(n)$ in terms of the cyclic permutations of length $1,2,..., n$.

We can thus write any term in the determinant as
\begin{equation}
M_{aa}\cdots M_{hh} \ (M_{ij} M_{jk} \cdots M_{ni}) \cdots (M_{pq} 
M_{qr} \cdots M_{zp})
\label{det.cycle}
\end{equation}
where there can be cyclic permutations of order $2,3,..., n$. 
Now, for each term of this form, there exists its adjoint where all 
the cycles have been reversed
\begin{equation}
M_{aa}\cdots M_{hh} \ (M_{ij} M_{jk} \cdots M_{ni})^{\rm R} \cdots 
(M_{pq} M_{qr} \cdots M_{zp})^{\rm R} \equiv M_{aa}\cdots M_{hh} \ 
(M_{in} \cdots M_{kj} M_{ji}) \cdots (M_{pz} \cdots M_{rq} M_{qp} )
\end{equation}

Let us first consider the simplest case where there is only one type 
of transition between any two states, that is the variable $\rho$ 
does not enter in the master equation. Any term $M_{ij}$
can then be written as 
\begin{equation}
M_{ij} =L_{ij}-s \, \delta_{ij}= \begin{cases} - \sum_k W(\omega_i\vert\omega_k) -s & \text{if $i=j$}\\
W(\omega_i\vert\omega_j) {\rm e}^{- z(\omega_i\vert\omega_j)} & \text{if $i\neq j$}
\end{cases}
\end{equation}
In this case, we can associate each term in the determinant with 
its adjoint and consider the quantity
\begin{equation}
M_{aa}\cdots M_{hh} \ (M_{ij} M_{jk} \cdots M_{ni}) \cdots (M_{pq} 
M_{qr} \cdots M_{zp}) +
M_{aa}\cdots M_{hh} \ (M_{ij} M_{jk} \cdots M_{ni})^{\rm R} \cdots 
(M_{pq} M_{qr} \cdots M_{zp})^{\rm R}
\label{L+Ladj}
\end{equation}
Introducing the notations $W_{aa} = - \sum_b W(\omega_a\vert\omega_b)$
and $W_{ij} = W(\omega_i\vert\omega_j)$, this term can be written as
\begin{equation}
(W_{aa}-s)\cdots (W_{hh}-s) \ (W_{ij} W_{jk} \cdots W_{ni}) \cdots 
(W_{pq} W_{qr} \cdots W_{zp}) \Big[ \prod_{{C}} \Big( 
\prod_{\omega\in{C}} {\rm e}^{- z(\omega\vert\omega')} \Big) + 
\prod_{{C}} \Big( {\rm e}^{- {\cal A}({C})} 
\prod_{\omega\in{C}} {\rm e}^{z(\omega\vert\omega')} \Big) \Big]
\label{det+adj}
\end{equation}
where the product is taken over all the cycles $\{{C}\}$
appearing in the decomposition (\ref{det.cycle}).
We also used that ${\rm e}^{-z(\omega\vert\omega)} = 1$, ${\rm 
e}^{-z(\omega\vert\omega')} = {\rm e}^{z(\omega'\vert\omega)}$ and 
equation (\ref{A.decomp}) for the affinity of a cycle so that
\begin{equation}
\frac{W_{ij} W_{jk} \cdots W_{ni}}{W_{in} \cdots W_{kj} W_{ji}} 
\equiv {\rm e}^{{\cal A}({C})}
\end{equation}
Moreover the quantity $\prod_{\omega\in{C}} {\rm e}^{- 
z(\omega\vert\omega')}$ can be written in terms of the scalar product 
(\ref{PS}) as
\begin{equation}
\prod_{\omega\in{C}} {\rm e}^{- z(\omega\vert\omega')}
= {\rm e}^{- \sum_l ({C},{C}_l)\lambda_l }
\end{equation}
Now, we can introduce this expression along with the expression (\ref{Adecom}) 
for the affinity in Eq. (\ref{det+adj}) to obtain
\begin{eqnarray}
& & (W_{aa}-s)\cdots (W_{hh}-s) \ (W_{ij} W_{jk} \cdots W_{ni}) 
\cdots (W_{pq} W_{qr} \cdots W_{zp}) \Big[ \prod_{{C}} {\rm e}^{- 
\sum_l ({C},{C}_l) \lambda_l} + \prod_{{C}} {\rm e}^{- 
\sum_l \ ({C},{C}_l) {\cal A}_l} {\rm e}^{\sum_l 
({C},{C}_l)\lambda_l} \Big] \nonumber \\
&=& (W_{aa}-s)\cdots (W_{hh}-s) \ (W_{ij} W_{jk} \cdots W_{ni}) 
\cdots (W_{pq} W_{qr} \cdots W_{zp}) \Big[ {\rm e}^{- \sum_l 
\sum_{{C}} ({C},{C}_l)\lambda_l } + {\rm e}^{\sum_l 
\sum_{{C}} ({C},{C}_l)\  ( \lambda_l - {\cal A}_l)} \Big]
\label{det.sym}
\end{eqnarray}
hence the symmetry
\begin{equation}
\lambda_l \rightarrow {\cal A}_l - \lambda_l \quad \textrm{for all}\ l
\label{transform}
\end{equation}
As it is valid for any term, the determinant also 
presents this symmetry. Consequently, the eigenvalues given by the 
roots of the characteristic equation will also be invariant under the 
transformation (\ref{transform}).

It is important for the preceding reasoning to be valid that a term 
and its adjoint occur with the same sign in the expression of the 
determinant. This is guaranteed because a permutation and its adjoint 
have the same structure in cycles (i.e., the same number of cycles of 
length $k$, for all $k$) hence a basic property of the symmetric 
groups ${\rm Sym}(n)$ implies that they are in the same conjugation 
class so that they have the same parity.

Now, we can consider the case where there is more than one mechanism 
of transition between two states. The reasoning is similar except 
that the terms $L_{ij}$ are now written as 
\begin{equation}
L_{ij}= \begin{cases} - \sum_{\rho} \sum_k W_{\rho}(\omega_i\vert\omega_k)  & \text{if $i=j$}\\
\sum_{\rho} W_{\rho}(\omega_i\vert\omega_j) {\rm e}^{- z_{\rho}(\omega_i\vert\omega_j)} & \text{if $i\neq j$}
\end{cases}
\end{equation}
A nontrivial cycle $(L_{ij} L_{jk} \cdots L_{ni})$ in 
the determinant is then written as
\begin{eqnarray}
(L_{ij} L_{jk} \cdots L_{ni}) &=& \Big[\sum_{\rho_1} 
W_{\rho_1}(\omega_i\vert\omega_j) {\rm e}^{- 
z_{\rho_1}(\omega_i\vert\omega_j)}\Big]\Big[\sum_{\rho_2} 
W_{\rho_2}(\omega_j\vert\omega_k) {\rm e}^{- 
z_{\rho_2}(\omega_j\vert\omega_k)}\Big] \cdots \Big[\sum_{\rho_r} 
W_{\rho_r}(\omega_n\vert\omega_i) {\rm e}^{- 
z_{\rho_r}(\omega_n\vert\omega_i)}\Big] \nonumber \\
&=& \sum_{\rho_1} \sum_{\rho_2} \cdots \sum_{\rho_r} 
W_{\rho_1}(\omega_i\vert\omega_j) {\rm e}^{- 
z_{\rho_1}(\omega_i\vert\omega_j)} W_{\rho_2}(\omega_j\vert\omega_k) 
{\rm e}^{- z_{\rho_2}(\omega_j\vert\omega_k)} \cdots 
W_{\rho_r}(\omega_n\vert\omega_i) {\rm e}^{- 
z_{\rho_r}(\omega_n\vert\omega_i)}
\end{eqnarray}
if the permutation is of order $r$. This is equivalent to summing 
over all the different possible cyclic paths
${\cal P}$ between the states $(\omega_i\omega_j\cdots \omega_n\omega_i)$:
\begin{equation}
(L_{ij} L_{jk} \cdots L_{ni}) = \sum_{\{ {\cal P} \}} (W_{ij} W_{jk} 
\cdots W_{ni})_{{\cal P}} \prod_{\omega,\rho\in{C}_{{\cal P}}} {\rm 
e}^{- z_{\rho}(\omega\vert\omega')}
\end{equation}
where $W_{ij}$ stands for $W_{\rho_1}(\omega_i\vert\omega_j)$ 
and the index ${\cal P}$ is a possible path between the states 
$(\omega_i\omega_j\cdots \omega_n\omega_i)$ obtained by selecting one of the 
possible transitions $\rho$ between these states:
\begin{equation}
{\cal P} = \omega_i \ {\overset{\rho_1}\longrightarrow} \ \omega_j 
{\overset{\rho_2}\longrightarrow} \
\omega_{2} {\overset{\rho_3}\longrightarrow} \cdots \omega_{n} 
{\overset{\rho_r}\longrightarrow} \ \omega_{i}
\end{equation}
Each of these possibilities also form a possible cycle hence we can 
make the same reasoning as before and associate a term with its 
adjoint. Its adjoint will also contain all possible paths between the 
states but in a reversed way so that the terms can be regrouped with 
each other to extract the affinity as in Eq. (\ref{det.sym}). Indeed, 
if there are $v$ cyclic permutations in the decomposition (\ref{det.cycle}),
the quantity (\ref{L+Ladj}) can here be expanded as
\begin{eqnarray}
(W_{aa}-s) & \cdots & (W_{hh}-s) \nonumber \\
 \times \sum_{\{ {\cal P}_1 \}} \cdots 
\sum_{\{ {\cal P}_v \}} & \Big\lbrace & \Big[ (W_{ij} W_{jk} \cdots 
W_{ni})_{{\cal P}_1} \prod_{\omega,\rho\in{C}_{{\cal P}_1}} {\rm e}^{- 
z_{\rho}(\omega\vert\omega')}\Big] \cdots\Big[ (W_{pq} W_{qr} \cdots 
W_{zp})_{{\cal P}_v} \prod_{\omega,\rho\in{C}_{{\cal P}_v}} {\rm e}^{- 
z_{\rho}(\omega\vert\omega')}\Big] \nonumber \\
&+& \Big[(W_{ij} W_{jk} \cdots W_{ni})^{\rm R}_{{\cal P}_1} 
\prod_{\omega,\rho\in{C}_{{\cal P}_1}} {\rm 
e}^{z_{\rho}(\omega\vert\omega')}\Big] \cdots \Big[(W_{pq} W_{qr} 
\cdots W_{zp})^{\rm R}_{{\cal P}_v} \prod_{\omega,\rho\in{C}_{{\cal 
P}_v}} {\rm e}^{z_{\rho}(\omega\vert\omega')} \Big] \; \Big\rbrace
\end{eqnarray}
with $W_{aa} = - \sum_{\rho} \sum_b W_{\rho}(\omega_a\vert\omega_b)$. 
We can put in evidence the transition rates to get
\begin{eqnarray}
& & (W_{aa}-s)\cdots (W_{hh}-s) \ \sum_{\{ {\cal P}_1 \}} \cdots 
\sum_{\{ {\cal P}_v \}} \Big\lbrace (W_{ij} W_{jk} \cdots 
W_{ni})_{{\cal P}_1} \cdots (W_{pq} W_{qr} \cdots W_{zp})_{{\cal 
P}_v}  \nonumber \\
&\times&  \Big[  \prod_{\omega,\rho\in{C}_{{\cal P}_1}} {\rm e}^{- 
z_{\rho}(\omega\vert\omega')} \cdots  \prod_{\omega,\rho\in{C}_{{\cal 
P}_v}} {\rm e}^{- z_{\rho}(\omega\vert\omega')} + {\rm e}^{- {\cal 
A}({C}_{{\cal P}_1})} \prod_{\omega,\rho\in{C}_{{\cal P}_1}} {\rm 
e}^{z_{\rho}(\omega_i\vert\omega_j)} \cdots {\rm e}^{- {\cal 
A}({C}_{{\cal P}_v})} \prod_{\omega,\rho\in{C}_{{\cal P}_v}} {\rm 
e}^{z_{\rho}(\omega_i\vert\omega_j)} \Big] \Big\rbrace
\end{eqnarray}
where we used the same relations as before. Decomposing the 
quantities $\prod_{\omega,\rho\in{C}_{{\cal P}}} {\rm e}^{\pm 
z_{\rho}(\omega_i\vert\omega_j)}$ and the affinities ${\cal 
A}({C}_{{\cal P}})$ in term of the fundamental set, we get
\begin{eqnarray}
& & (W_{aa}-s)\cdots (W_{hh}-s) \ \sum_{\{ {\cal P}_1 \}} \cdots 
\sum_{\{ {\cal P}_v \}}
(W_{ij} W_{jk} \cdots W_{ni})_{{\cal P}_1} \cdots (W_{pq} W_{qr} 
\cdots W_{zp})_{{\cal P}_v}  \nonumber \\
&\times&  \Big[ {\rm e}^{- \sum_l ({C}_{{\cal 
P}_1},{C}_l)\lambda_l} \cdots {\rm e}^{- \sum_l ({C}_{{\cal 
P}_k},{C}_l)\lambda_l} + {\rm e}^{\sum_l ({C}_{{\cal 
P}_1},{C}_l) (\lambda_l - {\cal A}_l)} \cdots {\rm e}^{\sum_l 
({C}_{{\cal P}_v},{C}_l) (\lambda_l - {\cal A}_l)} \Big] 
\nonumber \\
\end{eqnarray}
or equivalently
\begin{eqnarray}
(W_{aa}-s)\cdots (W_{hh}-s) \ \sum_{\{ {\cal P}_1 \}} \cdots \sum_{\{ 
{\cal P}_v \}} &&
(W_{ij} W_{jk} \cdots W_{ni})_{{\cal P}_1} \cdots (W_{ij} W_{jk} 
\cdots W_{ni})_{{\cal P}_v} \nonumber \\
&\times&  \Big[ {\rm e}^{- \sum_l \sum_{\nu}({C}_{{\cal 
P}_{\nu}},{C}_l)\lambda_l} + {\rm e}^{\sum_l \sum_{\nu} 
({C}_{{\cal P}_{\nu}},{C}_l)(\lambda_l - {\cal A}_l) } \Big]
\end{eqnarray}
The symmetry (\ref{transform}) is thus valid even in the more 
complicated case where there are several possible transitions between 
two states so that the corresponding characteristic equation also
presents the symmetry (\ref{transform}). The proof can be extended to  
Markovian processes on infinite state space either by using truncation
or by replacing the characteristic determinant by (\ref{ch.det}) by
a functional determinant, 
in which cases the symmetry is valid in the domain of convergence.

Finally, we can regroup the different mesoscopic currents associated
with the same macroscopic current $j_{\gamma}$
and which correspond the same affinity ${\cal A}_{\gamma} \equiv 
{\cal A}({C}_{\gamma})$, as discussed previously.
We thus set $\lambda_{l} = \lambda_{\gamma}$ for $l \in \gamma$ so that
\begin{equation}
j_{\gamma} (t) = \sum_{l \in \gamma} j_l (t)
\qquad \mbox{and}\qquad
G_{\gamma} (t) = \sum_{l \in \gamma} G_l (t)
\end{equation}
We thus obtain our central result:

\vskip 0.2 cm

{\bf Fluctuation theorem for the currents}: 
{\it Under the Schnakenberg conditions (\ref{ratio}), 
the generating function of the fluctuating integrated currents $G_{\gamma}(t)$ 
obeys the symmetry}
\begin{eqnarray}
Q(\{\lambda_{\gamma} \}; \{ {\cal A}_{\gamma} \}) =
\lim_{t\to\infty} -\frac{1}{t} \ln \langle {\rm e}^{- \sum_{\gamma} 
\lambda_{\gamma} G_{\gamma}(t)}\rangle
= Q(\{ {\cal A}_{\gamma} - \lambda_{\gamma} \}; \{ {\cal A}_{\gamma} \})
\label{FT.new.gen.macro}
\end{eqnarray}

\vskip 0.2 cm

In the nonequilibrium steady state, the mean value of the current 
$j_{\gamma} (t)$ is given by
\begin{equation}
{\cal J}_{\gamma} \equiv  \frac{\partial Q}{\partial
\lambda_{\gamma}}\Big\vert_{\{ \lambda_{\gamma}=0 \}}
= \lim_{t\to\infty} \frac{1}{t}\int_0^t dt' \langle j_{\gamma}(t')\rangle
= \lim_{t\to\infty} \frac{1}{t}\langle G_{\gamma}(t) \rangle
\label{q.macro.flux}
\end{equation}
Its higher-order moments can be obtained by successive derivations,
which shows that this function generates the statistical moments of 
the currents.
We notice that reversing the affinities corresponds to reversing
the sign of the currents and their fluctuations.

The generating function (\ref{FT.new.gen.macro}) is related to a 
large-deviation property of
the probability distribution of $(1/t) \int_{0}^{t} dt' j_{\gamma} 
(t')$ in the following way.
The decay rate of the probability that the currents averaged over a 
time interval $t$
take given values
\begin{equation}
H(\{ \alpha_{\gamma} \} ) \equiv
\lim_{t \rightarrow \infty} - \frac{1}{t} \ln \text{Prob}\left[ 
\left\{  \frac{1}{t} \int_{0}^{t} dt' j_{\gamma} (t') 
\in \left( \alpha_{\gamma},\alpha_{\gamma}+d\alpha_{\gamma}\right)\right\} \right]
\end{equation}
is the Legendre transform of the generating function:
\begin{equation}
H(\{ \alpha_{\gamma} \} )={\rm Max}_{\{\lambda_{\gamma}\}} \left[ 
Q(\{\lambda_{\gamma}\}) - \sum_{\gamma} 
\lambda_{\gamma}\alpha_{\gamma}\right]
\end{equation}
The symmetry relation (\ref{FT.new.gen.macro}) yields
\begin{equation}
H(\{- \alpha_{\gamma} \} ) - H(\{ \alpha_{\gamma} \} )  =
\sum_{\gamma} {\cal A}_{\gamma} \alpha_{\gamma}
\label{FT.new.tl}
\end{equation}
or equivalently
\begin{equation}
\frac{ \text{Prob}\left[\left\{ \frac{1}{t}\int_0^t dt' j_{\gamma}(t')
\in \left(\alpha_{\gamma},\alpha_{\gamma}+d\alpha_{\gamma}\right)
\right\} \right] }{\text{Prob}\left[\left\{ \frac{1}{t}\int_0^t dt'
j_{\gamma}(t') \in \left(-\alpha_{\gamma},-\alpha_{\gamma}+d\alpha_{\gamma}\right) \right\} \right] } 
\simeq \exp{ \sum_{\gamma}  {\cal A}_{\gamma} \alpha_{\gamma} t} 
\qquad\qquad (t \rightarrow \infty )
\label{CFTrap}
\end{equation}
which is the usual form of a fluctuation theorem.

We now have a fluctuation theorem for the macroscopic currents
which are physically measurable quantities.
We emphasize that, thanks to the above construction, this new 
symmetry in terms of the macroscopic affinities is completely general 
in the range of validity of Schnakenberg conditions (\ref{ratio}).
The present framework allows us to clearly identify the macroscopic currents
between the reservoirs maintaining the system out of equilibrium.

Another important aspect of this construction concerns the numerical 
obtention of the generating function (\ref{Q.flux}). Indeed if we 
integrate Eq. (\ref{G.eq}), the quantity $\sum_{\omega} 
F_t(\omega,\{\lambda_l\})$ will decay in the long-time limit as
\begin{equation}
\sum_{\omega} F_t(\omega,\{\lambda_l\}) \sim {\rm e}^{-Q(\{\lambda_l\})t}
\label{sum.G}
\end{equation}
The generating function can thus be obtained from the rate of the 
exponential decrease of this quantity. Such an algorithm is important 
because it does not strongly depend on the nonequilibrium conditions. 
In contrast, a direct statistical approach requires a very large 
statistics in order to obtain the generating function. This 
statistics grows exponentially fast as the parameters 
$\lambda_{\gamma}$ differ from zero
and as the nonequilibrium constraints increase.

\section{Connection with the Lebowitz-Spohn action functional}
\label{LSFT}

In  this section, we discuss the connection between the fluctuation 
theorem for the currents and the fluctuation theorem 
for the action functional by Lebowitz and Spohn \cite{LS99}.
These authors have introduced the action functional:
\begin{equation}
Z(t) \equiv \ln \frac{W_{\rho_1}(\omega_{0}\vert 
\omega_{1})W_{\rho_2}(\omega_{1}\vert \omega_{2}) \cdots
W_{\rho_n}(\omega_{n-1}\vert \omega_{n})}
{W_{\rho_1}(\omega_{1}\vert \omega_{0})W_{\rho_2}(\omega_{2}\vert \omega_{1})
\cdots W_{\rho_n}(\omega_{n}\vert \omega_{n-1})}
\label{Z}
\end{equation}
measuring the breaking of detailed balance along a random path
\begin{equation}
\omega (t) = \omega_{0} \ {\overset{\rho_1}\longrightarrow} \ 
\omega_{1} {\overset{\rho_2}\longrightarrow} \
\omega_{2}{\overset{\rho_3}\longrightarrow} \cdots 
{\overset{\rho_n}\longrightarrow} \ \omega_{n}
\label{path}
\end{equation}

The generating function of the fluctuating quantity (\ref{Z}) is defined by
\begin{equation}
q(\eta) \equiv \lim_{t\to\infty} -\frac{1}{t} \ln \langle {\rm 
e}^{-\eta Z(t)}\rangle
\label{Q}
\end{equation}
and presents the symmetry
\begin{equation}
q(\eta) = q(1-\eta)
\label{FT.Q}
\end{equation}
The mean value of the quantity (\ref{Z}) grows 
at a rate given by the entropy production in a 
nonequilibrium steady state \cite{LS99,G04,S05,AG04}:
\begin{equation}
 \frac{dq}{d\eta}(0) = \lim_{t\to\infty} 
\frac{\langle Z(t)\rangle}{t} = \frac{d_{\rm i}S}{dt}\Big\vert_{\rm st}
\label{Z.entroprod}
\end{equation}

The connection between the two fluctuation theorems 
(\ref{FT.new.gen.macro}) and (\ref{FT.Q})
can be established by the following reasoning.
The action functional (\ref{Z}) can be written as
\begin{equation}
Z(t) = \sum_e B_e \int_{0}^{t} dt' j_e(t')
\label{Zsep}
\end{equation}
in terms of the quantities (\ref{Be}).
We here use the convention that the orientation of the fundamental 
cycles are such that $S_l({C}_l)=1$.
This means that the cycles are oriented in the same way as the chords 
$l$ but this is not restrictive.

If we associate a different parameter $\eta_e$ with each possible 
transition, we can define the multi-variable generating function
\begin{equation}
{\cal Q}(\{ \eta_e \}) \equiv \lim_{t\to\infty} -\frac{1}{t} \ln 
\langle {\rm e}^{- \sum_e \eta_e B_e \int_0^{t} dt' j_e(t')}\rangle
\label{gen.Be}
\end{equation}
We recover the generating function $q(\eta)$ by setting $\eta_e = 
\eta$ for all the edges $e$ so that
\begin{equation}
q(\eta) = {\cal Q}(\{ \eta \})
\end{equation}

The generating function (\ref{gen.Be}) obeys the fluctuation theorem
\begin{equation}
{\cal Q}(\{ \eta_e \}) = {\cal Q}(\{ 1-\eta_e \})
\label{FT.eta_e}
\end{equation}
as proved here below.
We can absorb the local quantities $B_e$ to obtain the generating 
function of the currents on all the edges $e$ as
\begin{equation}
\tilde{{\cal Q}}(\{ \lambda_e \}) \equiv \lim_{t\to\infty} 
-\frac{1}{t} \ln \langle {\rm e}^{- \sum_e \lambda_e \int_0^{t} dt' 
j_e(t')}\rangle = {\cal Q}(\{ \lambda_e / B_e \})
\end{equation}
which should obey the symmetry
\begin{equation}
\tilde{{\cal Q}}(\{ \lambda_e \}) = \tilde{{\cal Q}} (\{ B_e - \lambda_e \})
\label{FT.Be}
\end{equation}
Indeed, the evolution operator for this generating function can be derived
as done in the previous section for the generating function of the 
independent currents.
The only difference is that there is a different parameter $\lambda$ 
for each possible transition, and
not only for the transitions corresponding to the chords $l$.
This amounts to redefine the quantities $z_\rho (\omega\vert\omega')$ as
\begin{equation}
{\rm e}^{- z_{\rho}(\omega \vert \omega')} \equiv \exp \Big[- \sum_e 
S_e (\omega \overset{\rho}{\rightarrow} \omega') \lambda_e \Big]
\end{equation}
where the sum is taken over all the edges $e$.
Let us consider the case where there is only one type of transition 
between two states.
The general case can be treated in a similar way. Using that
\begin{equation}
\prod_{\omega\in{C}} {\rm e}^{- z(\omega\vert\omega')} = {\rm 
e}^{- \sum_e S_e ({C})\lambda_e }
\end{equation}
a term in the characteristic determinant $\det\hat M=\det(\hat 
L-s\hat I)$ plus its adjoint can be written as
\begin{eqnarray}
& &M_{aa}\cdots M_{hh} \ (M_{ij} M_{jk} \cdots M_{ni}) \cdots (M_{pq} 
M_{qr} \cdots M_{zp}) +
M_{aa}\cdots M_{hh} \ (M_{ij} M_{jk} \cdots M_{ni})^{\rm R}
\cdots (M_{pq} M_{qr} \cdots M_{zp})^{\rm R}  \nonumber \\
&=& (W_{aa}-s)\cdots (W_{hh}-s) \ (W_{ij} W_{jk} \cdots W_{ni}) 
\cdots (W_{pq} W_{qr} \cdots W_{zp}) \Big[ \prod_{{C}} {\rm e}^{- 
\sum_e S_e({C})\lambda_e}
+ \prod_{{C}} {\rm e}^{- {\cal A}({C})} {\rm e}^{ \sum_e 
S_e({C})\lambda_e} \Big]
\label{calc.edge}
\end{eqnarray}
Using Eq. (\ref{A.decomp}), we can check that the transformation
\begin{equation}
\lambda_e \rightarrow B_e - \lambda_e
\label{symm.Be}
\end{equation} 
leaves this term invariant,
which proves the fluctuation theorem (\ref{FT.Be}) and thus (\ref{FT.eta_e}).

We notice that, if the chords $l$ are separated
from the other edges $e \not= \{l\}$, the transformation
\begin{eqnarray}
\lambda_l &\rightarrow& {\cal A}_l - \lambda_l \nonumber \\
\lambda_e &\rightarrow& - \lambda_e \quad (e\not= \{l\}) 
\label{symm.Al}
\end{eqnarray}
also leaves invariant the expression (\ref{calc.edge}).
The demonstration of the previous section can thus be applied and one 
finds that the generating function defined as
\begin{equation}
Q(\{\lambda_l \}) \equiv \tilde{{\cal Q}}(\{ \lambda_l \}, \{ 
\lambda_e=0 \}_{e\not=\{l\}})
\end{equation}
obeys the fluctuation theorem (\ref{CFT.Q}) because
\begin{eqnarray}
Q(\{\lambda_l \}) = \tilde{{\cal Q}}(\{ \lambda_l \}, \{ 
\lambda_e=0 \}_{e\not=\{l\}}) &=& \tilde{{\cal Q}}(\{ B_l - \lambda_l \}, \{ 
B_e \}_{e\not=\{l\}}) \nonumber \\
&=& \tilde{{\cal Q}}(\{ {\cal A}_l - \lambda_l \}, \{ 0 \}_{e\not=\{l\}})  = Q(\{{\cal A}_l -\lambda_l \})
\end{eqnarray}
where the second equality results from the symmetry (\ref{symm.Be})
and the third from the symmetry (\ref{symm.Al}).
Accordingly, the fluctuation theorem (\ref{FT.eta_e}) or (\ref{FT.Be}) implies 
\begin{equation}
Q(\{\lambda_{\gamma} \}) = Q(\{{\cal A}_{\gamma}-\lambda_{\gamma} \})
\end{equation}
which is the fluctuation theorem (\ref{FT.new.gen.macro}) 
for the macroscopic currents.
This establishes the link between the generating functions and their 
corresponding fluctuation theorem.  

Finally, the entropy production can be obtained from the measurement 
of the probability distribution of the currents and using the corresponding fluctuation theorem. 
Indeed, one can first notice that the mean value of the action functional $Z(t)$ can be 
expressed as the sum over all the mean currents
crossing the chords multiplied by the associated affinities.  
Precisely, thanks to equation (\ref{A.decomp}) and our choice of orientation, we 
can write the affinities of the fundamental set as
\begin{equation}
{\cal A}_l = \sum_e S_e({C}_l )B_e = B_l + \sum_{e\not= l} 
S_e({C}_l)B_e
\label{Asep}
\end{equation}
If we separate the chords from the other edges in the sum 
(\ref{Zsep}) and use Eq. (\ref{Asep}),
we get
\begin{eqnarray}
Z(t) &=& \sum_l  B_l \int_{0}^{t} dt' j_l(t') + \sum_{e \not= 
\{l\}}  B_e \int_{0}^{t} dt' j_e(t') \nonumber \\
&=& \sum_l  {\cal A}_l \int_{0}^{t} dt' j_l(t') - \sum_l 
\sum_{e\not= l} S_e({C}_l) B_e \int_{0}^{t} dt' j_l(t') + 
\sum_{e \not= \{l\}}  B_e \int_{0}^{t} dt' j_e(t') \nonumber \\
&=& \sum_l  {\cal A}_l \int_{0}^{t} dt' j_l(t') +R(t)
\end{eqnarray}
Using the property that
\begin{equation}
S_l ({C}_{l'}) = \delta_{l,l'}
\end{equation}
if $l$ and $l'$ are chords, the term $R(t)$ can be transformed as
\begin{eqnarray}
R(t) &=& - \sum_l \sum_{e\not= l} S_e({C}_l)  B_e 
\int_{0}^{t} dt' j_l(t') + \sum_{e \not= \{l\}}  B_e 
\int_{0}^{t} dt' j_e(t') \nonumber \\
&=& \sum_{e \not= \{l\}} B_e \Big[ - \sum_l S_e({C}_l)  
\int_{0}^{t} dt' j_l(t') +  \int_{0}^{t} dt' j_e(t') \Big]
\label{rest}
\end{eqnarray}
Now, Kirchhoff current law (\ref{KCL.S}) implies 
that the mean value of the rest (\ref{rest}) identically vanishes.
Since the mean value of the action functional $Z(t)$ increases with the rate of entropy 
production by Eq. (\ref{Z.entroprod}),
we find that the entropy production is given by the sum of the products
of the macroscopic affinities with the currents:
\begin{equation}
\frac{d_{\rm i}S}{dt}\Big\vert_{\rm st} = \sum_{\gamma} {\cal 
A}_\gamma {\cal J}_{\gamma}
\label{entropy.prod}
\end{equation}
By the fluctuation relation (\ref{CFTrap}), we can determine the 
affinities $\{{\cal A}_{\gamma}\}$ from the
measurement of the probability distribution of the currents, which 
also gives the mean currents
${\cal J}_{\gamma}$. The entropy production can thus be obtained 
thanks to the formula (\ref{entropy.prod})
solely in terms of the nonequilibrium fluctuations of the currents.


\section{Conclusions}
\label{Conclusions}

In this paper, we have established the fluctuation theorem for 
the currents for stochastic processes in which the macroscopic thermodynamic forces or affinities are identified thanks to the conditions (\ref{ratio}) of Schnakenberg \cite{S76}.  
This class of stochastic processes includes models of electronic transport
in mesoscopic conductors \cite{AWBMJ91,AG06}, biophysical models of ion transport
across membranes \cite{HK66,H68,H66}, and autocatalytic chemical reactions 
\cite{NP71,N72,MN75,NT77,NP77,JVN84,AG04}. In these stochastic processes,
the mesoscopic affinities (\ref{meso.affinities}), which are often considered,
depend on the mesoscopic state variables and are thus fluctuating in time,
as illustrated for the case of Schl\"ogl's reaction scheme in Subsection \ref{Example}. 
Instead, Schnakenberg network theory provides a decomposition of the graph
associated with the stochastic process in terms of cycles.
The ratio of the products of the transition states along each cycle and its time reversal
is directly related to a macroscopic thermodynamic force or affinity, as shown by
Schnakenberg \cite{S76}.  This allows us to define the macroscopic affinities
and the corresponding currents in the systems here considered.

In the proof of the fluctuation theorem for the currents, Schnakenberg's cycles 
have been identified in the characteristic determinant 
of the operator giving the generating function of the currents
as its leading eigenvalue. The cycles
of the graph naturally appear in the structure
of the determinant.  In this way, we can take the Schnakenberg conditions (\ref{ratio})
into account and prove the fluctuation theorem for the currents.
We have also proved that the fluctuation theorems
for the currents and for the Lebowitz-Spohn action functional 
can be deduced from another one for the generating function of
the mesoscopic currents on each edge of the graph.
This established a connection between the different fluctuation theorems,
which concern different fluctuating quantities.
Moreover, we have also shown that the fluctuation 
theorem and the measurement of the fluctuating currents should 
provide a new procedure to obtain the entropy production even if the 
transition rates are not known, as it is often the case in an 
experimental situation.

In conclusion, this theory can be applied to a large variety of processes in physics, 
chemistry, or biology.  The fluctuation theorem for the currents have important
consequences for nonequilibrium nanosystems, 
which we hope to report on in a future publication.

\vspace{0.3cm}

{\bf Acknowledgments.}
The authors thank Professor G.~Nicolis for support and encouragement in
this research. D.~Andrieux is Research Fellow at the F.N.R.S. Belgium.
This research is financially supported by the ``Communaut\'e fran\c 
caise de Belgique''
(contract ``Actions de Recherche Concert\'ees'' No. 04/09-312) and
the National Fund for Scientific Research (F.~N.~R.~S. Belgium, 
contract F. R. F. C. No. 2.4577.04).


\end{document}